# A Novel Ergodic Capacity Analysis of Diversity Combining and Multihop Transmission Systems over Generalized Composite Fading Channels[*]

Ferkan Yilmaz [1] and Mohamed-Slim Alouini [2]
*Electrical Engineering Program, Division of Physical Sciences and Engineering,*
*King Abdullah University of Science and Technology (KAUST),*
*Thuwal, Mekkah Province, Saudi Arabia.*
[1,2]{ferkan.yilmaz,slim.alouini}@kaust.edu.sa

*Abstract*—Ergodic capacity is an important performance measure associated with reliable communication at the highest rate at which information can be sent over the channel with a negligible probability of error. In the shadow of this definition, diversity receivers (such as selection combining, equal-gain combining and maximal-ratio combining) and transmission techniques (such as cascaded fading channels, amplify-and-forward multihop transmission) are deployed in mitigating various performance impairing effects such as fading and shadowing in digital radio communication links. However, the *exact analysis* of ergodic capacity is in general not always possible for all of these forms of diversity receivers and transmission techniques over generalized composite fading environments due to it's mathematical intractability. In the literature, published papers concerning the *exact analysis* of ergodic capacity have been therefore scarce (i.e., only [1] and [2]) when compared to those concerning the exact analysis of average symbol error probability. In addition, they are essentially targeting to the ergodic capacity of the maximal ratio combining diversity receivers and are not readily applicable to the capacity analysis of the other diversity combiners / transmission techniques. In this paper, we propose a novel moment generating function-based approach for the exact ergodic capacity analysis of both diversity receivers and transmission techniques over generalized composite fading environments. As such, we demonstrate how to simultaneously treat the ergodic capacity analysis of all forms of both diversity receivers and multihop transmission techniques.

## I. INTRODUCTION

Wireless systems continue to strive for higher data rates and better reliability and to migrate to higher and higher frequency bands. Due to high data-rate and coverage requirements of current systems, the ergodic capacity analysis of diversity receivers and transmission techniques becomes an important and fundamental issue from both theoretical and practical viewpoints. However and to the best of our knowledge, published papers concerning the *exact ergodic capacity analysis* of wireless communications systems have been scarce [1], [2] when compared to those concerning the exact symbol error probability (SEP) analysis. [3].

Shannon capacity of both diversity receivers and transmission techniques operating with bandwidth $W$ over generalized fading channels can be given in general by $C_{\gamma_{end}} \triangleq W \log_2(1+\gamma_{end})$, where $\gamma_{end}$ is the overall instantaneous SNR and $\log_2(\cdot)$ denotes the binary logarithm (i.e, it is the logarithm to the base 2). The ergodic capacity $C_{avg} \equiv \mathbb{E}\left[W \log_2\left(1+\gamma_{end}\right)\right]$, where $\mathbb{E}\left[\cdot\right]$ the expectation operator, is defined by averaging the Shannon capacity $C_{\gamma_{end}}$ over the probability density function (PDF) of the overall instantaneous SNR $\gamma_{end}$, i.e.,

$$C_{avg} = W \int_0^\infty \log_2\left(1+\gamma\right) p_{\gamma_{end}}(\gamma)\, d\gamma, \qquad (1)$$

where $p_{\gamma_{end}}(\gamma)$ is the PDF of the overall instantaneous SNR $\gamma_{end}$. If one has a closed form PDF for the $\gamma_{end}$, then (1) is very easy and straightforward to compute, yielding closed forms. On the other hand, important to note that $\gamma_{end}$ is in general a nonlinear function of the instantaneous SNRs $\gamma_1, \gamma_2, \ldots, \gamma_L$ such that $\gamma_{end} = \gamma_{end}(\gamma_1, \gamma_2, \ldots, \gamma_L)$, where the function $\gamma_{end}(\cdot)$ is the overall combining / transmitting function of the wireless communications systems. Diversity receivers and multihop transmission techniques have different nonlinear the overall combining / transmitting functions. Often and because of the nature of this nonlinearity, there is a high difficulty in the mathematical tractability of the statistical characterization of the overall instantaneous SNR $\gamma_{end}$. Hence, the PDF $p_{\gamma_{end}}(\gamma)$ of the overall instantaneous SNR $\gamma_{end}$ is generally not available in a simple and closed form and (1) has to be re-written as an $L$-fold integral in which the joint multivariate PDF of the instantaneous SNRs $\gamma_1, \gamma_2, \ldots, \gamma_L$, i.e., $p_{\gamma_1,\gamma_2,\ldots,\gamma_L}(r_1, r_2, \ldots, r_L)$ is needed.

Note that, even if under the assumption of that $p_{\gamma_1,\gamma_2,\ldots,\gamma_L}(r_1, r_2, \ldots, r_L)$ is somehow available in closed form, the $L$-fold integration is tedious and complicated in addition to the fact that it cannot be separated into a product

[*]This work has been presented by Ferkan Yilmaz in IEEE International Conference on Commnunications (ICC 2012), Ottawa, Canada, 10th-15th June, 2012. This work is presented to ensure a date and time stamp from the arXiv central authority "http://arxiv.org/" on the purpose of that this material is proven to be authors' scholarly and technical work. This work is moreover presented in the arXiv to share the authors' newly obtained results with the readership in advance of publication under the condition that copyright and all rights therein are retained by authors or by other copyright holders. All persons copying this work and accessing the information inside of it are strictly expected to adhere to the terms, conditions and constraints invoked by each author's. In most cases, this material cannot not be reposted, reannounced, reuploaded and reused without the explicit permission of the authors or the other copyright holders.
[*]This work was supported by King Abdullah University of Science and Technology (KAUST).

$$\mathcal{M}_{\gamma_{end}}(s) = \underbrace{\int_0^\infty \int_0^\infty \ldots \int_0^\infty}_{L\text{-fold}} \exp\left(-s\left(\sum_{\ell=1}^L \sqrt{r_\ell/L}\right)^2\right) p_{\gamma_1,\gamma_2,\ldots,\gamma_L}(r_1, r_2, \ldots, r_L)\, dr_1\, dr_2 \ldots dr_L. \qquad (4)$$

of one dimensional integrals. In addition, it is computationally cumbersome, especially as the number of branches / hops (i.e., $L$) increases. Thus, researchers in literature have tried to find the PDF of the overall instantaneous SNR $\gamma_{end}$ in order to find the ergodic capacity of the diversity receiver or the transmission technique they considered. Nevertheless, this technique is often complicated and tedious for generalized fading environment since it involves also an $L$-fold integral ($L$-fold convolution) even if the instantaneous SNRs $\gamma_1, \gamma_2, \ldots, \gamma_L$ are assumed to be independent.

For these reasons, two different MGF-based approaches have been *elegantly developed only* in [1], [2]. In particular, using [4, Eq. (3.434/2)], Hamdi developed in [1] an elegant MGF-based approach for the exact capacity analysis, i.e.,

$$C_{avg} = \frac{W}{\log(2)} \int_0^\infty \frac{e^{-s}}{s} \left[1 - \mathcal{M}_{\gamma_{end}}(s)\right] ds, \qquad (2)$$

where $\mathcal{M}_{\gamma_{end}}(s)$ is the MGF of the overall instantaneous SNR $\gamma_{end}$. Later, using [4, Eq. (6.224/1)], an alternative elegant MGF-based approach for the exact capacity analysis was developed by Di Renzo *et. al* in [2], i.e.,

$$C_{avg} = \frac{W}{\log(2)} \int_0^\infty \text{Ei}(-s) \left[\frac{\partial}{\partial s} \mathcal{M}_{\gamma_{end}}(s)\right] ds, \qquad (3)$$

where $\text{Ei}(\cdot)$ denotes the exponential integral function [4, Eq. (8.211/1)]. It is in this context important to note that the elegant MGF-based approaches developed in [1], [2] are computationally efficient and easy to use if and only if the overall MGF $\mathcal{M}_{\gamma_{end}}(s)$ can be written in closed-form. It is nevertheless not always the case. For example, for EGC diversity receiver whose overall instantaneous SNR $\gamma_{end}$ can be written as $\gamma_{end} = \left(\sum_{\ell=1}^L \sqrt{\gamma_\ell/L}\right)^2$, the overall MGF $\mathcal{M}_{\gamma_{end}}(s)$ can be given as in (4) at the top of this page. Referring to the main idea behind the MGF approach [3], an MGF-based analysis approach is computationally efficient and easy to use if and only if the overall MGF $\mathcal{M}_{\gamma_{end}}(s)$ can be written in the product of the MGFs of the branches / hops for the considered wireless communication systems, (4) can neither be obtained in closed-form nor be expressed as the product of the MGFs of the diversity branches even if the instantaneous SNRs $\gamma_1, \gamma_2, \ldots, \gamma_L$ are assumed to be mutually independent. As a consequence, the elegant MGF-based approaches developed in [1] and [2] are essentially targeting to the ergodic capacity analysis of maximal ratio combining (MRC) diversity receiver and are not applicable and extendible to the computation of the ergodic capacity of the other diversity receivers and transmission techniques.

In this paper, we propose a novel MGF-based approach for the ergodic capacity analysis over correlated / uncorrelated generalized composite fading channels and for an arbitrary number of diversity branches / hops. As such, in contradiction to both [1] and [2], using the MGF-based approach proposed in this paper demonstrates how to simultaneously treat the ergodic capacity analysis for a variety of diversity receivers (such as selection combining (SC), EGC, MRC and root-mean-square combining (RMSC)), transmission techniques (such as cascaded fading channels and amplify-and-forward (AF) multihop transmission) and some approximation techniques applied to overall instantaneous SNR $\gamma_{end}$ (such as geometric-mean approximation and minimum-bound SNR).

The remainder of this paper is organized as follows. In Section II, a generic expression for the overall instantaneous SNR $\gamma_{end}$ is intriguingly offered to unify the instantaneous SNR at the output of the diversity receivers and transmission techniques. Then, using this generic overall instantaneous SNR $\gamma_{end}$, we propose a novel MGF-based approach for the ergodic capacity analysis which is remarkably unified not only for a variety of diversity receivers and transmission techniques but also for generalized composite fading environments. In Section III, some important diversity receivers and transmission techniques are outlined and discussed In order to check analytical simplicity and accuracy of the proposed MGF-based approach and show how to simultaneously treat the ergodic capacity analysis for a variety of diversity receivers and transmission techniques. Finally, conclusions are drawn in the last section. Last but not the least, we obtained in Appendix the generalized MGF of the instantaneous SNR in the EGK fading environment, It should be mentioned that several proofs of our results have intentionally been omitted due to space limitation. However, all these results have been checked numerically for their correctness and accuracy.

## II. Novel Ergodic Capacity Analysis Unified for Diversity Receivers and Transmission Techniques

In both diversity receivers and transmission techniques, the overall instantaneous SNR $\gamma_{end}$ is a function of $\gamma_1, \gamma_2, \ldots, \gamma_L$ such that $\gamma_{end} = \gamma_{end}(\gamma_1, \gamma_2, \ldots, \gamma_L)$ as mentioned before. Explicitly shown in the following subsection that, in the largest majority of cases, the overall instantaneous SNR $\gamma_{end}$ can be put in a single and unified mean / aggregation form of the instantaneous SNRs $\gamma_1, \gamma_2, \ldots, \gamma_L$.

### A. Unified Overall Instantaneous SNR $\gamma_{end}$

The overall instantaneous SNR $\gamma_{end}$ (that is, it is the overall instantaneous SNR at the output of wireless communications systems such as diversity receivers and transmission techniques) can in general be consistently written as

$$\gamma_{end} = \gamma_{end}(\eta, p, q) = \eta \left(\frac{1}{L} \sum_{\ell=1}^L \gamma_\ell^p\right)^q, \qquad (5)$$

where $\eta \in \mathbb{R}^+$, $p \in \mathbb{Z}^+$ and $q \in \mathbb{Z}^+$ specify the type of the combining or transmission technique (see Table I at the top of the next page). Note that $\eta$ denotes the power-scaling

TABLE I
SOME SPECIAL CASES OF THE UNIFIED OVERALL INSTANTANEOUS SNR

**MRC Receiver** has the overall instantaneous SNR [3]:
$$\gamma_{mrc} = \gamma_{end}(L, 1, 1) = \sum_{\ell=1}^{L} \gamma_\ell.$$

**EGC Receiver** has the overall instantaneous SNR [3]:
$$\gamma_{egc} = \gamma_{end}(L, 1/2, 2) = \frac{1}{L}\left(\sum_{\ell=1}^{L} \gamma_\ell^{\frac{1}{2}}\right)^2.$$

**SC Receiver** has the overall instantaneous SNR [3]:
$$\gamma_{sc} = \lim_{p\to\infty} \gamma_{end}(1, p, 1/p) = \max(\gamma_1, \gamma_2, \ldots, \gamma_L).$$

**RMSC Receiver** has the overall instantaneous SNR [5], [6]:
$$\gamma_{rmsc} = \gamma_{end}(\sqrt{L}, 2, 1/2) = \sqrt{\sum_{\ell=1}^{L} \gamma_\ell^2}.$$

**Cascaded Fading Channel** has the overall instantaneous SNR [7, and the references therein]:
$$\gamma_{cc} = \lim_{p\to 0} \gamma_{end}(1, pL, 1/p) = \prod_{\ell=1}^{L} \gamma_\ell.$$

**Geometric-Mean Approximation of the Overall instantaneous SNR** can be written as [8], [9]
$$\gamma_{gm} = \lim_{p\to 0} \gamma_{end}(1, p, 1/p) = \sqrt[L]{\prod_{\ell=1}^{L} \gamma_\ell}.$$

**AF Multihop Transmission** has the overall instantaneous SNR [10, and the references therein]
$$\gamma_{mh} = \gamma_{end}(1/L, -1, -1) = \frac{1}{\frac{1}{\gamma_1} + \frac{1}{\gamma_2} + \cdots + \frac{1}{\gamma_L}}.$$

**Minimum-Bound of the Overall instantaneous SNR** can be written as
$$\gamma_{min} = \lim_{p\to-\infty} \gamma_{end}(1, p, 1/p) = \min(\gamma_1, \gamma_2, \ldots, \gamma_L).$$

factor of the $L$-branch diversity receiver / $L$-hop transmission technique (such as $\eta = L$ for both MRC and EGC receivers, and $\eta = 1$ for SC receiver, and $\eta = 1/L$ for AF multihop transmission). Furthermore, for the transmitted signal with the average symbol energy $E_s$, the instantaneous SNR $\gamma_\ell$ of the $\ell$th branch / hop is written as $\gamma_\ell = \frac{E_s}{N_0}\alpha_\ell^2$, where $\alpha_\ell$ and $N_0$ denote the amplitude of the fading and the power of the additive white Gaussian noise (AWGN), respectively, of the $\ell$th branch / hop.

It is of interest to note that, referring to both (1) and (5), the ergodic capacity analysis, which is remarkably unified for a variety of diversity receivers and transmission techniques, can in general be written as

$$C_{avg} = W \underbrace{\int_0^\infty \int_0^\infty \ldots \int_0^\infty}_{L\text{-fold}} \log_2\left(1 + \eta\left(\frac{1}{L}\sum_{\ell=1}^{L} r_\ell^p\right)^q\right) \times$$
$$p_{\gamma_1,\gamma_2,\ldots,\gamma_L}(r_1, r_2, \ldots, r_L) dr_1 dr_2 \ldots dr_L. \quad (6)$$

where the $L$-fold integration in (6) is tedious and complicated in addition to the fact that it cannot be separated into a product of one dimensional integrals.

In what follows, we present a new exact and *unified* MGF-based approach that overcomes the difficulty mentioned above. As such, referring to special cases of (5), we offer a generic and unified single integral expression to simultaneously treat the ergodic capacity of diversity receivers and transmission techniques over generalized composite fading channels.

### B. An MGF-based Unification of the Average Capacity Analysis in Generalized Fading Channels

The unified analysis to be explained in this subsection allows for the capacity analysis of wireless communications systems characterized by a large variety of the combinations of combining techniques and transmission techniques in generalized composite fading environments.

**Theorem 1** (Unified and Generic Ergodic Capacity in Correlated and Generalized Composite Fading Environments). *Upon being unified and generic for a variety of diversity receivers and transmission techniques, the ergodic capacity in correlated and generalized composite fading environments is given by*[1]

$$\mathcal{C}_{avg} = \frac{W}{\log(2)} \int_0^\infty C_{\eta,q}(s)\left[\frac{\partial}{\partial s}\mathcal{M}_{\gamma_1^p, \gamma_2^p, \ldots, \gamma_L^p}(s)\right] ds, \quad (7)$$

*where $\eta \in \mathbb{R}^+$, $p \in \mathbb{R}$ and $q \in \mathbb{R}$ are chosen according to the type of the combining or transmission technique, and where $C_{\eta,q}(s)$ is the auxiliary function given by*

$$C_{\eta,q}(s) = -H_{3,3}^{1,2}\left[\frac{\eta}{L^q s^q} \;\middle|\; \begin{array}{l}(1,1),(1,1),(1,\frac{|q|+q}{2})\\(1,1),(0,1),(0,\frac{|q|-q}{2})\end{array}\right], \quad (8)$$

*where $H_{p,q}^{m,n}[\cdot]$ is the Fox's H function [12, Eq. (8.3.1/1)].*[2] *Moreover, $\mathcal{M}_{\gamma_1^p, \gamma_2^p, \ldots, \gamma_L^p}(s) \equiv \mathbb{E}[\exp(-s\sum_\ell \gamma_\ell^p)]$ is the joint generalized MGF of the instantaneous SNRs $\gamma_1, \gamma_2, \ldots, \gamma_L$ of the branches / hops.*

*Proof:* The proof is omitted due to page limitation ∎

In order to find the ergodic capacity, the novel MGF-based approach proposed in Theorem 1 eliminates the necessity of finding the PDF of the overall instantaneous SNR $\gamma_{end}$ and show how to obtain the ergodic capacity unified and generic for a variety of diversity receivers and transmission techniques, by means of conspicuously converting $L$-fold integration into a single integration using the joint generalized MGF $\mathcal{M}_{\gamma_1^p, \gamma_2^p, \ldots, \gamma_L^p}(s)$. In addition in the case of there is no correlation between all instantaneous SNRs $\gamma_1, \gamma_2, \ldots, \gamma_L$, the unified and generic ergodic capacity can be given by the following corollary.

**Corollary 1** (Unified and Generic Ergodic Capacity in Mutually Independent and Non-Identically Distributed Fading Environments). *The exact average capacity analysis in mutually independent and non-identically distributed fading channels*

---
[1]Note that the two integrals in Theorem 1 and Corollary 1 can be easily computed by standard mathematical software packages such as Mathematica®, Matlab® and Maple™ or they can be readily and accurately estimated by employing the Gauss-Chebyshev quadrature (GCQ) formula [11, Eq.(25.4.39)], which converges rapidly and steadily, requiring few terms for an accurate result.

[2]For more information about the Fox's H function, the readers are referred to [13], [14]

*with the bandwidth $W$ is given by*[1]

$$\mathcal{C}_{avg} = \frac{W}{\log(2)} \int_0^\infty \mathrm{C}_{\eta,q}(s) \sum_{\ell=1}^{L} \left[\frac{\partial}{\partial s}\mathcal{M}_{\gamma_\ell^p}(s)\right] \prod_{\substack{k=1 \\ k\neq\ell}}^{L} \mathcal{M}_{\gamma_k^p}(s) ds \quad (9)$$

*where, for $\ell \in \{1, 2, \ldots, L\}$, $\mathcal{M}_{\gamma_\ell^p}(s) \equiv \mathbb{E}\left[\exp\left(-s\gamma_\ell^p\right)\right]$ is the generalized MGF $\mathcal{M}_{\gamma_\ell^p}(s)$ of the instantaneous SNR $\gamma_\ell$ corresponding to the $\ell$th branch / hop.*

*Proof:* The proof is omitted due to page limitation. ∎

Note that, in order to evaluate Corollary 1, the generalized MGF $\mathcal{M}_{\gamma_\ell^p}(s)$ of the instantaneous SNR $\gamma_\ell$ corresponding to the $\ell$th branch / hop and its derivative are obviously needed and given in closed-form for the extended generalized-K (EGK) fading environments which is a more generalized composite fading distribution whose special cases includes many more well-known fading distributions (i.e., see Appendix). In this context, useful to mention that, regarding the numerical computation of (8), an efficient Mathematica® implementation of the Fox's H function is available in [7, Appendix A]. Note moreover that the auxiliary function $\mathrm{C}_{\eta,q}(s)$ can also be expressed in terms of more familiar Meijer's G function (which is for example available as a built-in function in Mathematica®), using [12, Eq. (8.3.2/22)]. More specifically, the Meijer's G representation of (8) can be found for the rational integer values of the parameter $q$ (that is, if we restrict $|q|$ to $|q| = k/l$, where $k$ and $l$ are arbitrary positive integers) as shown in the following corollary.

**Corollary 2** (Meijer's G Representation of the Auxiliary Function $\mathrm{C}_{\eta,q}(s)$). *The auxiliary function $\mathrm{C}_q(s)$ can be given in terms of the more familiar Meijer's G function for the rational integer values of the parameter $q$, i.e., when $q$ is restricted to $|q| = k/l$, where $k$ and $l$ are arbitrary positive integers. In this case, the auxiliary function $\mathrm{C}_{\eta,q}(s)$ is given by*

$$\mathrm{C}_{\eta,q}(s) = -\sqrt{\frac{(2\pi)^{k+1}}{(2\pi)^{2l}k}} \mathrm{G}_{2l+k,2l}^{l,2l}\left[\frac{\eta^l k^k}{L^k s^k} \left| \begin{array}{c} \Xi_{(1)}^{(l)}, \Xi_{(1)}^{(l)}, \Xi_{(1)}^{(k)} \\ \Xi_{(1)}^{(l)}, \Xi_{(0)}^{(l)} \end{array}\right.\right], \quad (10a)$$

*for $q \geq 0$;*

$$\mathrm{C}_{\eta,q}(s) = -\sqrt{\frac{(2\pi)^{k+1}}{(2\pi)^{2l}k}} \mathrm{G}_{2l,2l+k}^{l,2l}\left[\frac{\eta^l L^k s^k}{k^k} \left| \begin{array}{c} \Xi_{(1)}^{(l)}, \Xi_{(1)}^{(l)} \\ \Xi_{(1)}^{(l)}, \Xi_{(0)}^{(l)}, \Xi_{(0)}^{(k)} \end{array}\right.\right]. \quad (10b)$$

*for $q < 0$, where $\Xi_{(n)}^{(x)} \equiv \frac{x}{n}, \frac{x+1}{n}, \frac{x+2}{n}, \ldots, \frac{x+n-1}{n}$ with $x \in \mathbb{C}$ and $n \in \mathbb{Z}^+$.*

Note that the number of coefficients of the Meijer's G function in both (10a) and (10b) increases as the values of both $k \in \mathbb{Z}^+$ and $l \in \mathbb{Z}^+$ increase regarding $|q| = k/l$, so much so that the computation complexity of (15b) considerably grows and its corresponding computation latency substantially increases. Therefore, $k \in \mathbb{Z}^+$ and $l \in \mathbb{Z}^+$ should be kept as much as smaller while supporting the condition $|q| = k/l$. More specifically, the computation efficiency and latency of Meijer's G function $\mathrm{G}_{p,q}^{m,n}[\cdot]$ is primarily addressed by the total number of coefficients (i.e., $p + q$). In this context, the complexity of both (10a) and (10b) can be seen as proportionated to their total number of coefficients $(4l + k)$. As such, the utilization of the Fox's H function implementation in [7, Appendix A] in these instances is preferable.

### III. SPECIAL CASES

Despite the fact that the novel results in the previous section are easy to use, let us consider in this section the special cases of the auxiliary function $\mathrm{C}_{\eta,q}(s)$ (see Table I) in order to check analytical simplicity and accuracy of how the novel MGF-based approach presented in Theorem 1 and Corollary 1 simultaneously treats the ergodic capacity analysis of diversity receivers and transmission techniques over generalized composite fading channels.

**Special Case 1** (MRC Diversity Receiver). Referring to the unified auxiliary function given by (8), the parameters $\eta$, $p$ and $q$ for the $L$-branch MRC diversity receiver are set to $\eta = L$, $p = 1$ and $q = 1$, respectively. Then, using [12, Eqs. (8.3.2/21) and (8.2.2/9)], the auxiliary function $\mathrm{C}_{L,1}(s)$ can be expressed as

$$\mathrm{C}_{L,1}(s) = -\mathrm{H}_{3,2}^{1,2}\left[\frac{1}{s} \left| \begin{array}{c} (1,1), (1,1), (1,1) \\ (1,1), (0,1) \end{array}\right.\right], \quad (11a)$$

$$= -\mathrm{G}_{2,1}^{0,2}\left[\frac{1}{s} \left| \begin{array}{c} 1, 1 \\ 0 \end{array}\right.\right]. \quad (11b)$$

As a step forward, utilizing [12, Eq. (8.2.2/14)] and [12, Eq. (8.4.11/1)] together, one can readily reduce (11b) into

$$\mathrm{C}_{L,1}(s) = \mathrm{Ei}(-s), \quad (12)$$

where $\mathrm{Ei}(\cdot)$ is the exponential integral function [4, Eq. (8.211/1)]. In order to check the validity, substituting (12) into (7) results in the well-known ergodic capacity result given in [2, Eq. (7)] by Di Renzo et. al. As such, substituting (12) into (9) further simplifies to the ergodic capacity $\mathcal{C}_{avg}$ of the MRC diversity receiver over mutually independent fading channels as expected. □

**Special Case 2** (EGC Diversity Receiver). Referring the special case of Table I with $\eta = L$, $p = 1/2$ and $q = 2$, and then using some algebraic manipulations based on [12, Eqs. (8.3.2/21) and (8.2.2/9)], the auxiliary function $\mathrm{C}_{L,2}(s)$ for the $L$-branch EGC diversity receiver can be readily given by

$$\mathrm{C}_{L,2}(s) = -\mathrm{H}_{3,2}^{1,2}\left[\frac{1}{L s^2} \left| \begin{array}{c} (1,1), (1,1), (1,2) \\ (1,1), (0,1) \end{array}\right.\right], \quad (13a)$$

$$= -\sqrt{\pi} \mathrm{G}_{3,1}^{0,2}\left[\frac{4}{Ls^2} \left| \begin{array}{c} 1, 1, \frac{1}{2} \\ 0 \end{array}\right.\right]. \quad (13b)$$

where using [12, Eq. (8.4.12/4)], $\mathrm{C}_{L,2}(s)$ results in

$$\mathrm{C}_{L,2}(s) = 2\,\mathrm{Ci}\left(\sqrt{L}s\right), \quad (14)$$

where $\mathrm{Ci}(x)$ is the cosine integral function [11, Eq. (5.2.27)]. Eventually, substituting (14) into (7) in order to check the validity,, the ergodic capacity $\mathcal{C}_{avg}$ of the EGC diversity receiver over correlated diversity branches results in [15, Eq. (4)] as expected. For the case that there does not exist any correlation among diversity branches, it is further simplified to [15, Eq. (5)] by means of substituting (14) into (9). □

**Special Case 3** (SC Diversity Receiver). Referring the special case of Table I with $\eta = 1$, $p \to \infty$ and $q = 1/p \to 0$ such that the parameter $p$ is chosen as an integer ($p \in \mathbb{Z}^+$) and $pq = 1$, the auxiliary function $C_{\eta,q}(s)$ given by (8) becomes

$$C_{\eta,\frac{1}{p}}(s) = -H_{3,2}^{1,2}\left[\frac{1}{\sqrt[p]{sL}} \,\middle|\, \begin{array}{c}(1,1),(1,1),(1,\frac{1}{p})\\(1,1),(0,1)\end{array}\right], \quad (15a)$$

$$= \frac{-1}{(2\pi)^{p-1}} G_{2p+1,2p}^{l,2p}\left[\frac{1}{Ls} \,\middle|\, \begin{array}{c}\Xi_{(1)}^{(p)}, \Xi_{(1)}^{(p)}, 1\\\Xi_{(p)}^{(l)}, \Xi_{(0)}^{(p)}\end{array}\right]. \quad (15b)$$

Finally, the ergodic capacity $C_{avg}$ of the $L$-branch SC diversity receiver having correlated branches is attained by substituting (15b) into (7). Moreover, (15b) into (9) results in the average capacity of the SC diversity receiver having uncorrelated branches. At this point, important to mention that the average performance of the SC diversity receiver is commonly computed in the literature using a cumulative distribution function (CDF)-based approach due to the fact that the CDF of the maximum of all instantaneous SNRs $\gamma_1, \gamma_2, \ldots, \gamma_L$ can be obtained as the product of the CDFs of all instantaneous SNRs in independent fading conditions. In contrast and to the best of our knowledge, the ergodic capacity analysis for the SC diversity receiver, which is presented above, proposes an new and novel MGF-based approach which was unknown in the literature. □

**Special Case 4** (RMSC Diversity Receiver). For RMSC diversity receiver [5], [6], the parameters $\eta$, $p$ and $q$ are set to $\eta = \sqrt{L}$, $p = 2$ and $q = 1/2$, respectively. Then, using (8), the auxiliary function $C_{\sqrt{L},\frac{1}{2}}(s)$ can be expressed as

$$C_{\sqrt{L},\frac{1}{2}}(s) = -H_{3,2}^{1,2}\left[\frac{1}{s^{\frac{1}{2}}} \,\middle|\, \begin{array}{c}(1,1),(1,1),(1,\frac{1}{2})\\(1,1),(0,1)\end{array}\right], \quad (16a)$$

$$= -\frac{1}{2\pi} G_{3,2}^{1,3}\left[\frac{1}{s} \,\middle|\, \begin{array}{c}\frac{1}{2},1,1\\\frac{1}{2},0\end{array}\right]. \quad (16b)$$

Using [16, Eq. (07.34.03.0528.01) and (07.25.03.0076.01)] together, the auxiliary function $C_{\sqrt{L},\frac{1}{2}}(s)$ given by (11b) reduces to

$$C_{\sqrt{L},\frac{1}{2}}(s) = \frac{1}{2}\left(\text{Ei}(s) - \sqrt{\frac{8s}{\pi}}\,_2F_2\left[\frac{1}{2},1;\frac{3}{2},\frac{3}{2};u\right]\right), \quad (17)$$

where $_2F_2[\cdot;\cdot;\cdot]$ is the hypergeometric function defined in [12, Eq. (7.2.3/1)]. With this result, the ergodic capacity $C_{avg}$ of the $L$-branch RMSC diversity receiver can be expressed using (7) in general. Furthermore, substituting (17) into (9) results in the ergodic capacity of the RMSC diversity receiver over mutually independent branches. □

**Special Case 5** (Cascaded Fading Channel). Referring the special case of Table I with $\eta = 1$, $q \to \infty$ and $p = L/q \to 0$ such that the parameter $p$ is chosen as an integer ($p \in \mathbb{Z}^+$) and $pq = L$, and using [12, Eq. (8.3.2/21)], the auxiliary function $C_{1,q}(s)$ for the $L$-hop cascaded fading channel can be readily given as

$$C_{1,q}(s) = -H_{3,2}^{1,2}\left[\frac{1}{L^q s^q} \,\middle|\, \begin{array}{c}(1,1),(1,1),(1,q)\\(1,1),(0,1)\end{array}\right], \quad (18a)$$

$$= -\sqrt{(2\pi)^{q-1}q^{-1}}\, G_{2+q,2}^{1,2}\left[\frac{q^q}{L^q s^q} \,\middle|\, \begin{array}{c}1,1,\Xi_{(1)}^{(q)}\\1,0\end{array}\right]. (18b)$$

where $q \in \mathbb{Z}^+$ is numerically chosen as high as possible (i.e., $q \gg 1$). Finally, substituting (18) into (7) or (9) results in the average capacity $C_{avg}$ of the correlated cascaded fading channel or the mutually independent cascaded fading channel. □

**Special Case 6** (Geometric-Mean Approximation of the Overall instantaneous SNR). After setting the parameters $\eta,p,q$ to $\eta = 1$, $q \to \infty$ and $p = \to 0$ so that $pq = 1$, the auxiliary function $C_{1,q}(s)$ for the geometric-mean approximation of the overall instantaneous SNR $\gamma_{end}$ can be readily obtained as (18a) and (18), in which $q \in \mathbb{Z}^+$ is numerically chosen as high as possible (i.e., $q \gg 1$). Eventually, the average capacity $C_{avg}$ for geometric-mean approximation of the overall instantaneous SNR $\gamma_{end}$ can be attained by utilizing both (7) or (9). □

**Special Case 7** (AF Multihop Transmission). For $L$-hop AF multihop transmission, the parameters $\eta$, $p$ and $q$ are set to $\eta = 1/L$, $p = -1$ and $q = -1$, respectively. Then, referring to (8), the auxiliary function $C_{\frac{1}{L},-1}(s)$ can be readily written as

$$C_{1/L,-1}(s) = -H_{2,3}^{1,2}\left[s \,\middle|\, \begin{array}{c}(1,1),(1,1)\\(1,1),(0,1),(0,1)\end{array}\right], \quad (19a)$$

$$= -G_{2,3}^{1,2}\left[s \,\middle|\, \begin{array}{c}1,1\\1,0,0\end{array}\right]. \quad (19b)$$

For the further simplification, (19b) reduces, by utilizing [16, Eq. (07.34.03.0475.01)] and [16, Eq. (06.06.03.0003.01)], into

$$C_{1/L,-1}(s) = \text{Ei}(-s) - \log(s) - \mathbf{C}, \quad (20)$$

where $\mathbf{C}$ is the Euler-Mascheroni constant (also called Euler's constant) [4, Eq.(8.367/1)]. Inserting (20) into (9), the average capacity of the $L$-hop AF multihop transmission can be readily obtained. It is interesting to point out that using the integration by parts rule [4, Eq.(2.02/5)], the resulting formula can be readily reduced into [10, Eq.(10)] as expected. □

**Special Case 8** (Minimum-Bound of the Overall instantaneous SNR). Referring the special case of Table I with $\eta = 1$, $p \to -\infty$ and $q = 1/p \to 0^-$ with $pq = 1$, and using (8), the auxiliary function $C_{1,1/p}(s)$ for the minimum-bound of the overall instantaneous SNR can be attained as

$$C_{\eta,\frac{1}{p}}(s) = -H_{2,3}^{1,2}\left[L^{\frac{1}{|p|}}s^{\frac{1}{|p|}} \,\middle|\, \begin{array}{c}(1,1),(1,1)\\(1,1),(0,1),(0,\frac{1}{|p|})\end{array}\right], \quad (21a)$$

$$= \frac{-1}{(2\pi)^{|p|-1}} G_{2|p|,2|p|+1}^{|p|,2|p|}\left[Ls \,\middle|\, \begin{array}{c}\Xi_{(1)}^{(|p|)},\Xi_{(1)}^{(|p|)}\\\Xi_{(1)}^{(|p|)},\Xi_{(0)}^{(|p|)},0\end{array}\right]. \quad (21b)$$

where $|p| \in \mathbb{Z}^+$ is numerically chosen as high as possible (i.e., $|p| \gg 1$). Finally, substituting (21b) into (9) results in the minimum-bound for the average capacity. □

Consequently, these all special cases prove the analytical accuracy and validity of how the novel MGF-based approach simultaneously treat the ergodic capacity analysis for a variety of diversity receivers and transmission techniques.

## IV. Conclusion

In this paper, a new MGF-based approach for the ergodic capacity analysis of diversity receivers and transmission techniques over generalized composite fading channels is presented. In contrast to the elegant MGF-based approaches available in the literature [1], [2] which are basically unified with respect to generalized fading channels, the novel MGF-based approach proposed in this paper is expressly more generic enough to unify the ergodic capacity analysis for popular diversity receivers (such as selection combining, maximal ratio combining, equal gain combining and root-mean square combining) and transmission techniques (such as amplify-and-forward multihop transmission, cascaded fading channel, geometric-mean approximation of the overall instantaneous SNR $\gamma_{end}$ and minimum-bound approximation of the overall instantaneous SNR $\gamma_{end}$), i.e., there is no need to separately analyze the ergodic capacity of these diversity receivers and transmission techniques. As a consequence, note that the special cases demonstrated in Section III openly evidence and undoubtedly prove the methodological soundness, analytical accuracy, and validity of the unified and generic MGF-based ergodic analysis. Therefore, and due to space limitation, several proofs of our results have intentionally been omitted . However, all these results have been checked numerically for their correctness and accuracy. In addition, obviously combined with the fact the novel MGF based ergodic capacity analysis is obviously valid for a variety of generalized composite fading channels, the ergodic capacity analysis is demonstrated with the extended generalized-K (EGK) fading distribution.

## Appendix

The extended generalized-K (EGK) distribution is a generalized distribution that provides a unified theory to model the envelope statistics of known wireless and optical communication channels as explained in [17], [18]. More precisely, note that several distributions such as Rayleigh, lognormal, Weibull, Nakagami-$m$, generalized Nakagami-$m$, generalized-K and which are listed in [17, Table I], [18, Table I] are the special or limiting cases of EGK distribution. Regarding this disclosed versatility, the EGK distribution offers a unified theory as to statistically model the envelope statistics of known wireless / optical communication channels. In EGK fading channels, the PDF for the instantaneous SNR $\gamma_\ell$ of the $\ell$th branch / hop is given by [17, Eq.(3)], [18, Eq.(5)]

$$p_{\gamma_\ell}(\gamma) = \frac{\xi}{\Gamma(m_s)\Gamma(m)} \left(\frac{\beta_s \beta}{\bar{\gamma}}\right)^{m\xi} \gamma^{m\xi-1} \times \Gamma\left(m_s - m\frac{\xi}{\xi_s}, 0, \left(\frac{\beta_s \beta}{\bar{\gamma}}\right)^{m\xi} \gamma^\xi, \frac{\xi}{\xi_s}\right), \quad (A.1)$$

where $m$ ($0.5 \leq m < \infty$) and $\xi$ ($0 \leq \xi < \infty$) represent the fading figure (diversity severity / order) and the fading shaping factor, respectively, while $m_s$ ($0.5 \leq m_s < \infty$) and $\xi_s$ ($0 \leq \xi < \infty$) represent the shadowing severity and the shadowing shaping factor (inhomogeneity), respectively. In addition, the parameters $\beta$ and $\beta_s$ are defined as $\beta = \Gamma(m+1/\xi)/\Gamma(m)$ and $\beta_s = \Gamma(m_s + 1/\xi_s)/\Gamma(m_s)$, respectively, where $\Gamma(\cdot)$ is the Gamma function [11, Eq. (6.5.3)]. Furthermore, in (A.1), $\Gamma(\cdot,\cdot,\cdot,\cdot)$ is the extended incomplete Gamma function defined as $\Gamma(\alpha, x, b, \beta) = \int_x^\infty r^{\alpha-1} \exp\left(-r - br^{-\beta}\right) dr$, where $\alpha, \beta, b \in \mathbb{C}$ and $x \in \mathbb{R}^+$ [19, Eq. (6.2)].

After some algebraic manipulations, the unified generalized MGF $\mathcal{M}_{\gamma_\ell^p}(s) \triangleq \mathbb{E}\left[\exp(-s\gamma_\ell^p)\right]$ can be given by

$$\mathcal{M}_{\gamma_\ell^p}(s) = \frac{1}{\Gamma(m)\Gamma(m^s)|p|} \times H^{3,1}_{1,3}\left[\frac{\beta\beta_s}{\bar{\gamma}s^{1/p}} \middle| \begin{array}{c} (1, \phi(p)) \\ (m, \frac{1}{\xi}), (m_s, \frac{1}{\xi_s}), (0, \phi(-p)) \end{array}\right] \quad (A.2)$$

where the coefficient function $\phi(p)$ is given as $\phi(p) = \mathrm{U}(p)/|p| + \mathrm{U}(-p)$, where $\mathrm{U}(\cdot)$ denotes the unit step function [4], [20]. Moreover, its derivative can be given by

$$\frac{\partial}{\partial s}\mathcal{M}_{\gamma_\ell^p}(s) = \frac{1}{\Gamma(m)\Gamma(m^s)|p|p\,s} \times H^{4,1}_{2,4}\left[\frac{\beta\beta_s}{\bar{\gamma}s^{1/p}} \middle| \begin{array}{c} (1, \phi(p)), (0,1) \\ (1,1), (m, \frac{1}{\xi}), (m_s, \frac{1}{\xi_s}), (0, \phi(-p)) \end{array}\right]. \quad (A.3)$$

Finally, substituting (A.2) and (A.3) into (9), the unified and generic MGF-based ergodic capacity analysis can be readily obtained.

## References

[1] K. A. Hamdi, "Capacity of MRC on correlated Rician fading channels," vol. 56, no. 5, pp. 708–711, May 2008.
[2] M. Di Renzo, F. Graziosi, and F. Santucci, "Channel capacity over generalized fading channels: A novel MGF-based approach for performance analysis and design of wireless communication systems," *IEEE Transactions on Vehicular Tech.*, vol. 59, no. 1, pp. 127–149, Jan. 2010.
[3] M. K. Simon and M. S. Alouini, *Digital Communication over Fading Channels*, 2nd ed. Hoboken, New Jersey. USA: John Wiley & Sons, Inc., 2005.
[4] I. S. Gradshteyn and I. M. Ryzhik, *Table of Integrals, Series, and Products*, 5th ed. San Diego, CA: Academic Press, 1994.
[5] O. Abdul-Latif and J.-P. Dubois, "LS-SVM detector for RMSGC diversity in SIMO channels," in *The 9th International Symposium on Signal Processing and Its Applications (ISSPA)*, Feb. 2007, pp. 1–4.
[6] ——, "LS-SVM detector for RMSGC diversity in SIMO channels," in *inproceeding of International Symposium on Signal Processing and Its Applications (ISSPA 2007)*, Feb. 2007, pp. 1–4.
[7] F. Yilmaz and M.-S. Alouini, "Product of the powers of generalized Nakagami-m variates and performance of cascaded fading channels," in *IEEE Global Communications Conference (GLOBECOM 2009), Honolulu, Hawaii, USA*, Nov. 30-Dec. 4 2009.
[8] N. C. Sagias, G. K. Karagiannidis, P. T. Mathiopoulos, and T. A. Tsiftsis, "On the performance analysis of equal-gain diversity receivers over generalized Gamma fading channels," *IEEE Transactions on Wireless Communications*, vol. 5, no. 10, pp. 2967–2975, Oct. 2006.
[9] G. K. Karagiannidis, "Performance bounds of multihop wireless communications with blind relays over generalized fading channels," vol. 5, no. 3, pp. 498–503, Mar. 2006.
[10] F. Yilmaz, O. Kucur, and M.-S. Alouini, "Exact capacity analysis of multihop transmission over amplify-and-forward relay fading channels," in *IEEE Inter. Symp. on Personal Indoor and Mobile Radio Commun. (PIMRC) - Istanbul, Turkey*, Sep. 2010, pp. 2293–2298.
[11] M. Abramowitz and I. A. Stegun, *Handbook of Mathematical Functions with Formulas, Graphs, and Mathematical Tables*, 9th ed. New York, USA: Dover Publications, 1972.
[12] A. P. Prudnikov, Y. A. Brychkov, and O. I. Marichev, *Integral and Series: Volume 3, More Special Functions*. CRC Press Inc., 1990.
[13] A. Kilbas and M. Saigo, *H-Transforms: Theory and Applications*. Boca Raton, Florida, USA: CRC Press LLC, 2004.

[14] A. M. Mathai, R. K. Saxena, and H. J. Haubold, *The H-Function: Theory and Applications*, 1st ed.   Dordrecht, Heidelberg, London, New York: Springer Science, 2009.
[15] F. Yilmaz and M.-S. Alouini, "An MGF-based capacity analysis of equal gain combining over fading channels," in *proceeding of IEEE International Symposium on Personal Indoor and Mobile Radio Communications (PIMRC 2010), Istanbul, Turkey*, Sep. 2010, pp. 945–950.
[16] Wolfram Research, *Mathematica Edition: Version 8.0*.   Champaign, Illinois: Wolfram Research, Inc., 2010.
[17] F. Yilmaz and M.-S. Alouini, "A new simple model for composite fading channels: Second order statistics and channel capacity," in *International Symposium on Wireless Communication Systems (ISWCS 2010) - York, UK*, Sep. 2010, pp. 676–680.
[18] ——, "Extended generalized-K (EGK): A new simple and general model for composite fading channels," *Submitted to IEEE Transactions on Communications*, Dec. 12 2010, Available at http://arxiv.org/abs/1012.2598.
[19] M. A. Chaudhry and S. M. Zubair, *On a Class of Incomplete Gamma Functions with Applications*, 1st ed.   Boca Raton-London-New York-Washington, D.C.: Chapman & Hall/CRC, 2002.
[20] D. Zwillinger, *CRC Standard Mathematical Tables and Formulae*, 31st ed.   Boca Raton, FL: Chapman & Hall/CRC, 2003.